\documentclass[twocolumn,preprintnumbers,amsmath,amssymb,superscriptaddress,longbibliography, prl]{revtex4-1}

\usepackage{graphicx}
\usepackage{bm}
\usepackage{dsfont}
\usepackage[usenames,dvipsnames]{xcolor}
\usepackage{pstricks}
\usepackage[tight]{subfigure}
\usepackage{verbatim}
\usepackage{units}
\usepackage{multirow}
\usepackage{enumitem}
\usepackage{mathrsfs}
\usepackage{leftidx}
\usepackage{xspace}
\usepackage{braket}
\usepackage{bbm}
\usepackage{cancel}
\usepackage{amsfonts,amssymb,amsmath}
\usepackage[normalem]{ulem}

\usepackage[a4paper]{hyperref}
\hypersetup{colorlinks=true,linktoc=all,linkcolor=blue,breaklinks=true,citecolor=blue,urlcolor=blue}
\usepackage[english]{babel}
\renewcommand{\L}{_\text{c}}
\newcommand{\R}{_\text{h}}

\newcommand{\fL}{f_\text{c}(E)}
\newcommand{\fR}{f_\text{h}(E)}
\newcommand{\fa}{f_\alpha(E)}

\newcommand{\ThL}{\Theta_\text{c}}
\newcommand{\ThR}{\Theta_\text{h}}
\newcommand{\Ssh}{S_\text{sh}}

\newcommand{\kB}{k_\text{B}}

\newcommand{\cut}{\!\!\!}

\begin{document}

\title{Out-of-Equilibrium Fluctuation-Dissipation Bounds}

\author{Ludovico Tesser}
	\affiliation{Department of Microtechnology and Nanoscience (MC2), Chalmers University of Technology, S-412 96 G\"oteborg, Sweden}
	
	\author{Janine Splettstoesser}
	\affiliation{Department of Microtechnology and Nanoscience (MC2), Chalmers University of Technology, S-412 96 G\"oteborg, Sweden}
	
\date{\today}

\begin{abstract}


We prove a general inequality between the charge current and its fluctuations valid for any noninteracting coherent electronic conductor and for any stationary out-of-equilibrium condition, thereby going beyond established fluctuation-dissipation relations.
The developed  \emph{fluctuation-dissipation bound} saturates at large temperature bias and reveals additional insight for heat engines, since it limits the output power by power fluctuations. 
It is valid when the thermodynamic uncertainty relations break down due to quantum effects and provides stronger constraints close to thermovoltage.
\end{abstract}

\maketitle

Fluctuations (noise) in mesoscopic devices are attracting ever increasing interest~\cite{Blanter2000Sep,Esposito2009Dec} because of their role in the performance of  nanoscale heat engines~\cite{Benenti2017Jun,Pekola2021Oct} in addition to their opportunities for transport spectroscopy.
Indeed, on the one hand, noise measurements are instrumental for characterizing nanoscale systems~\cite{Blanter2000Sep}, e.g. revealing information about the temperature~\cite{White1996Aug} and temperature biases~\cite{Lumbroso2018Oct, Larocque2020Sep}, or detecting fractional charges~\cite{De-Picciotto1997Sep, Saminadayar1997Sep}. On the other hand, fluctuations limit the precision of device operations.
A pivotal result in the study of noise is the fluctuation-dissipation theorem (FDT)~\cite{Callen1951Jul, Green1954Mar, Kubo1957Jun}, establishing a fundamental connection between fluctuations and dissipation in systems at thermal equilibrium.
Beyond equilibrium the FDT does in general not hold and more complex features in fluctuations can appear. This is however an important situation, not only for steady-state heat engines, where large voltage and temperature biases allow to explore the intriguing nonlinear properties of nanoscale devices to boost their performance~\cite{Benenti2017Jun}.
The development of fluctuation relations~\cite{Esposito2009Dec}, which entail properties of the nonequilibrium transport, paved the way to further investigations on the out-of-equilibrium noise in the Markovian regime~\cite{Prost2009Aug, Baiesi2009Jul, Seifert2010Jan, Altaner2016OCt}, in weakly time-dependently driven systems~\cite{Moskalets2009Aug, Riwar2021Jan} or in ac or dc driven systems in the tunneling regime~\cite{Safi2020Jul, Safi2022Nov}. An important extension of the fluctuation-dissipation theorem in the context of mesoscopic electronic conductors is known, but only in the regime where tunneling between conductors is weak and for zero temperature bias~\cite{Levitov2004Sep}. 

In this Letter, we release these constraints and establish general out-of-equilibrium bounds on the  noise with respect to average charge currents. Importantly, they are derived without requiring local detailed balance to hold. 
Our bounds are valid for arbitrary noninteracting mesoscopic conductors under arbitrary stationary out-of-equilibrium conditions. This means in particular, that strong coupling between different contacts is included and that the contacts can be subject to large potential and temperature biases.
The inequalities that we develop set both upper and lower bounds on the fluctuations for given nonequilibrium conditions and for the average currents they induce. We refer to them as out-of-equilibrium fluctuation-dissipation bounds (FDB).
Previously, \textit{inequalities} constraining fluctuations have relied on the classical limit~\cite{Dechant2020Mar} or on local detailed balance. They have important implications for the operation of a device, such as the celebrated thermodynamic uncertainty relations (TUR)~\cite{Seifert2018Aug,Barato2015Apr,Timpanaro2019Aug, Hasegawa2019Sep,Saryal2019Oct,Horowitz2020Jan,Ehrlich2021Jul,Strasberg2022Book} constraining the signal-to-noise ratio of 
currents or output power by entropy production. 
They thereby offer insights into the performance limitations of heat engines at the nanoscale~\cite{Pietzonka2018May, Kheradsoud2019Aug, Holubec2021Dec}. 
Importantly, these TURs can be violated in coherent conductors beyond local detailed balance~\cite{Brandner2018Mar, Agarwalla2018Oct, Liu2019Jun, Saryal2019Oct, Timpanaro2021Jun, Gerry2022Apr}.  
By employing the out-of-equilibrium FDBs that we develop here, for devices operating as thermoelectric steady-state heat engines, we find a constraint between the output power and its fluctuations which is valid even when the TUR is not, and which provides a stronger constraint close to the thermovoltage, i.e. close to the finite voltage bias at which the produced power vanishes.

In order to develop these fluctuations-dissipation bounds, we study coherent and noninteracting electron transport in the steady-state using scattering theory~\cite{Blanter2000Sep, Moskalets2011Sep}.
For simplicity, here we focus on a two-terminal setup. Details of the discussion of the multi-terminal case are presented in the supplemental material~\cite{supp}.
Within this framework, the average charge current is
\begin{equation}
    I = \frac{q}{h}\int dE D(E)[\fL-\fR]
\end{equation}
where $q$ is the electron charge, $h$ is the Planck constant, and $D(E)\in[0,1]$ is the transmission probability of the conductor at energy $E$.
The Fermi distribution $f_\alpha (E)\equiv \{1+\exp[\beta_\alpha(E-\mu_\alpha)]\}^{-1}$ describes the electronic occupation of the colder and hotter reservoir $\alpha=\mathrm{c,h}$ with inverse temperature $\beta_\alpha \equiv 1/(\kB T_\alpha)$ and electrochemical potential $\mu_\alpha$.
We define the biases as $\Delta\mu\equiv \mu_\mathrm{h}-\mu_\mathrm{c}$ and $\Delta T\equiv T_\mathrm{h}-T_\mathrm{c}\geq0$ and the average temperature $\bar{T}=\left(T_\mathrm{h}+T_\mathrm{c}\right)/2$.
The zero-frequency noise of the charge current is defined as 
$S^I \equiv \int \langle\delta\hat{I}(t)\delta\hat{I}(0)\rangle dt $, where $\delta \hat{I} \equiv \hat{I}- I$ is the variation with respect to the average.
We separate the total noise as $S^I = \ThL^I + \ThR^I + \Ssh^I$ with
\begin{subequations}\label{eq:Noise-definitions}
\begin{align}
\Theta_\alpha^I & \equiv \frac{q^2}{h}\!\int_{-\infty}^\infty\cut\cut dE D(E) \fa[1-\fa],  \\
\Ssh^I & \equiv \frac{q^2}{h}\!\int_{-\infty}^\infty \cut\cut dE D(E)[1-D(E)] [\fL-\fR]^2.
\end{align}
\label{eq:noise-definitions}    
\end{subequations}
The thermal noise component $\Theta_\alpha^I$ depends only on reservoir $\alpha$, it vanishes when the temperature is zero and can be finite even at equilibrium, entailing the fluctuation dissipation theorem (FDT).
By contrast, the shot noise component $\Ssh^I$ vanishes at equilibrium $f\L=f\R$, and contains the partitioning factor $D(E)[1-D(E)]$. It hence vanishes when the outcome of the electron transmission is certain, $D(E)\in\{0,1\}$.
While in noise measurements the total noise $S^I$ is observed, it is possible to extract the thermal and the shot noise components by combining multiple noise measurements~\cite{Hasegawa2021Jan, Tesser2023Feb}.

An instance of the fluctuation-dissipation theorem is the Johnson-Nyquist noise~\cite{Johnson1927Jan, Nyquist1928Jul}, given by
   $S^I = -2q\kB \bar{T} \left.\partial I/\partial \Delta\mu\right|_{\Delta\mu\equiv0}$, 
with the conductance $-\partial I /\partial[\Delta\mu/q]|_{\Delta\mu\equiv0}$. Crucially, this result holds at equilibrium only, namely in the absence of temperature and electrochemical potential bias, i.e. $\Delta T = 0 = \Delta\mu$.
However, the Johnson-Nyquist relation has been extended to the presence of a finite electrochemical potential bias ($\Delta\mu\neq0$), while maintaining a zero temperature bias ($\Delta T=0$).
This generalization~\cite{Levitov2004Sep} relates the current fluctuation to the average current and to a $\coth$-factor that has its origin in the detailed-balance relation between tunneling rates, 
\begin{equation}\label{eq:FDT}
    S^I = q I \coth\left(\frac{\Delta\mu}{2\kB T}\right).
\end{equation}
It holds in the tunneling regime, which within scattering theory corresponds to $D(E)\ll 1$. Note that, for noninteracting systems, namely when Eqs.~(\ref{eq:Noise-definitions}) hold, we find an equivalent equation for the noise components and the current~\cite{supp}, which is valid in the high-transmission regime $[1-D(E)]\ll 1$,
\begin{equation}\label{eq:FDT2}
    \Ssh^I - \left(\ThL^I + \ThR^I\right)= \left(qI + \frac{q^2 \Delta\mu}{h}\right) \coth\left(\frac{\Delta \mu}{2\kB T}\right) - \frac{2q^2 \kB T}{h}.
\end{equation}
It complements Eq.~\eqref{eq:FDT} by its validity range and quantifies the \textit{difference} between shot and thermal noise.

\textit{Fluctuation-dissipation bounds.---}
Relaxing the previous constraints on transmission probabilities and nonequilibrium conditions, we now allow the transmission $D(E)$ to be arbitrary, and the temperature bias $\Delta T$ to be finite. 
A relevant quantity to characterize the out-of-equilibrium fluctuations is the (positive or negative) \textit{excess noise}, $S^I- 2\ThR^I$, namely the difference between the total noise and the thermal noise of the equilibrium setup at the hotter temperature and at reference potential $\mu_\mathrm{h}$.
To characterize the excess noise, we introduce in analogy to Ref.~\cite{Levitov2004Sep} the excess rates between the out-of-equilibrium and the equilibrium condition,
\begin{subequations}\label{eq:excess-rates}
    \begin{align}
        \tilde\Gamma_{\rightarrow} \equiv \!\!\int\!\!\frac{dE}{h}D(E)[\fL[1-\fR] - \fR[1-\fR]], \\
        \tilde\Gamma_{\leftarrow} \equiv \!\!\int\!\!\frac{dE}{h}D(E)[\fR[1-\fL] - \fR[1-\fR]].
    \end{align}
\end{subequations}
The average charge current is proportional to the difference between these excess rates, 
\begin{equation}\label{eq:current-with-excess-rates}
    I = q(\tilde\Gamma_\rightarrow - \tilde\Gamma_\leftarrow),
\end{equation}
where the equilibrium contribution to the rates cancels out.
By contrast, the excess noise is bounded by the sum of the excess rates, 
\begin{equation}\label{eq:fluctuation-dissipation-bound}
   S^I - 2\ThR^I \leq q^2(\tilde\Gamma_\rightarrow + \tilde\Gamma_\leftarrow) \leq -qI\tanh\left(\frac{\Delta\mu}{2\kB\Delta T}\right).
\end{equation}
This out-of-equilibrium fluctuation dissipation bound (FDB) is the central result of this Letter, based on which several important relations will be developed in the following. The first inequality of Eq.~\eqref{eq:fluctuation-dissipation-bound} relies on $D(1-D)\leq D$ and it is saturated in the tunneling regime, i.e. $D(E)\ll1$, where only independent single-particle tunnelings contribute. If, in addition to weak tunneling, also the temperatures in the two contacts are equal, $\Delta T=0$, it hence reduces to the statement of~\cite{Levitov2004Sep}, see Eq.~\eqref{eq:FDT}.
The second inequality exploits properties of the Fermi functions. It is saturated when the temperature bias $\kB\Delta T$ is the \emph{largest} energy scale, such that the excitations of the hot contact are all approximately equally occupied, see~\cite{supp} for details of the derivation.
This shows that the fluctuation-dissipation bound in Eq.~\eqref{eq:fluctuation-dissipation-bound} is fundamentally different from the generalized FDT in Eq.~\eqref{eq:FDT} because the latter requires $\kB\Delta T$ to be the \emph{smallest} energy scale in order to hold.
For large temperature bias, $\kB\Delta T\gg |\Delta\mu|$, we can approximate the right-hand side of Eq.~\eqref{eq:fluctuation-dissipation-bound} as
\begin{equation}\label{eq:FDB-approximation}
    S^I - 2\ThR^I \leq -\frac{q^2}{2\kB}\frac{P}{\Delta T}, \quad \mathrm{for} \quad \kB\Delta T\gg |\Delta\mu|
\end{equation}
where we introduced the output power
 $P\equiv I\Delta \mu/q$.
Equation~\eqref{eq:FDB-approximation} shows that the excess noise is limited by the output power. Specifically, if the device \textit{produces} power, i.e. $P>0$, the out-of-equilibrium noise $S^I$ is bounded to be smaller than the (hot) equilibrium noise $2\ThR^I$. By contrast, if the device \textit{dissipates} power, i.e. $P<0$, the out-of-equilibrium noise $S^I$ can exceed the (hot) equilibrium noise $2\ThR^I$ by at most a factor proportional to the dissipated power. When $P<0$, inequality~\eqref{eq:FDB-approximation} holds true for any temperature bias but is less constraining than~\eqref{eq:fluctuation-dissipation-bound}.

Note that the bound of Eq.~\eqref{eq:fluctuation-dissipation-bound} implies the zero-current shot noise bound that was recently established in Refs.~\cite{Eriksson2021Sep, Tesser2023Feb}. Indeed when setting $I=0$ in Eq.~\eqref{eq:fluctuation-dissipation-bound}, we find---using the noise decomposition in Eq.~\eqref{eq:Noise-definitions}---that the shot noise at zero current is bounded by the thermal noise difference, $\Ssh^I\leq \ThR^I - \ThL^I$.

Starting from Eq.~\eqref{eq:fluctuation-dissipation-bound}, an analogous fluctuation-dissipation bound can be derived for the excess noise with respect to the equilibrium noise of the system at the colder temperature, $S^I- 2\ThL^I$, by replacing $D(E)\rightarrow\tilde{D}(E) = 1- D(E)$. This bound 
\begin{equation}\label{eq:Noise-Upper-Bound}
    S^I - 2 \ThL^I \leq\frac{q^2 \kB}{h}\Delta T+ q\left(I+\frac{q\Delta\mu}{h}\right)\tanh\left(\frac{\Delta\mu}{2\kB\Delta T}\right),
\end{equation}
is tight in the high-transmission regime, i.e. when $[1-D(E)]\ll1$. 

Similar fluctuation-dissipation bounds to Eqs.~\eqref{eq:fluctuation-dissipation-bound},\eqref{eq:Noise-Upper-Bound} can even be developed for multi-terminal conductors with multiple channels  $N_\alpha$. Such a multi-terminal formulation even allows for including more general out-of-equilibrium situations such as spin-biases. We find the bound~\cite{supp}
\begin{widetext}
\begin{equation}\label{eq:FDB_general}
    S_{\text{hh}}^I - S_{\text{hh}, \text{eq}}^I \leq -\frac{q^2}{h}\sum_{\alpha\neq\text{h}}\tanh\left(\frac12\frac{\mu_\text{h}-\mu_\alpha}{\kB(T_\text{h}-T_\alpha)}\right)\int dE \sum_{i=1}^{N_\text{h}}D_{\text{h}\alpha, i}(E)\left[f_\alpha(E) - f_\text{h}(E)\right],
\end{equation}
\end{widetext}
where the total autocorrelation noise $S_{\text{hh}}^I\equiv \int \langle\delta\hat{I}_\text{h}(t)\delta\hat{I}_\text{h}(0)\rangle dt$ in the \textit{hottest} terminal is compared to the equilibrium noise $S_\text{hh,eq}^I$, namely the thermal noise obtained when all reservoirs have electrochemical potential $\mu_\text{h}$ and temperature $T_\text{h}$. The function $D_{\text{h}\alpha,i}(E)$ quantifies the transmission probability from contact $\alpha$ to h via the eigenchannel $i$. Similar to the multi-terminal generalization of the FDT of ~\eqref{eq:FDT}, see Ref.~\cite{Levitov2004Sep}, the generalized multi-terminal fluctuation-dissipation bound of Eq.~\eqref{eq:FDB_general} does not feature the average currents, but weighted  current contributions between different terminals.

We now use the FDB of Eq.~\eqref{eq:fluctuation-dissipation-bound} to find constraints on the total out-of-equilibrium current fluctuations, $S^I$.
Specifically, by exploiting that the noise components are non-negative $\Theta^I_\alpha, \Ssh^I\geq0$, and that the thermal noise satisfies $\Theta^I_\alpha \leq q^2 \kB T_\alpha/h$, we find
\begin{widetext}
\begin{align}
    &qI \tanh\left(\frac{\Delta\mu}{2\kB \Delta T}\right) \leq S^I \leq \frac{2q^2 }{h}\kB\bar{T} + \left(qI+\frac{q^2 \Delta\mu}{h}\right)\tanh\left(\frac{\Delta\mu}{2\kB \Delta T}\right). 
\label{eq:S-tot-bounds}
\end{align}
\end{widetext}
These constraints on the total noise $S^I$ establish a relation between the average current $I$ and its fluctuations $S^I$ at given nonequilibrium conditions, valid for \emph{any} out-of-equilibrium condition and noninteracting mesoscopic conductors with \emph{any} transmission $D(E)$.

\begin{figure}
    \centering
    \includegraphics{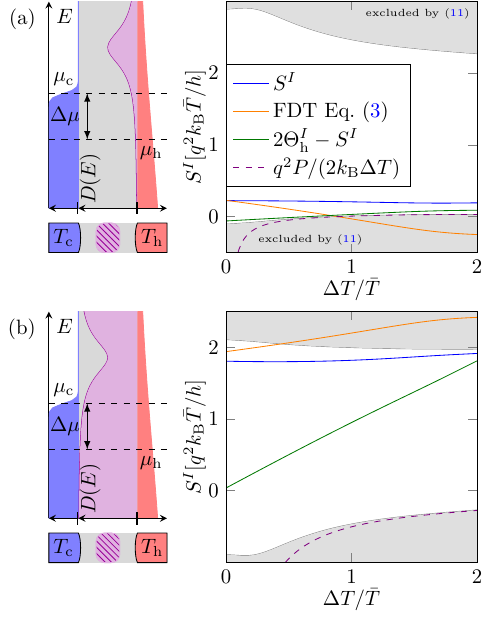}
    \caption{Noise and its constraints for Lorentzian $D_\text{Lor}(E)$ (a) and anti-Lorentzian $[1-D_\text{Lor}(E)]$ (b) transmissions. The average temperature $\Bar{T}$ is kept constant and used to evaluate the FDT expression in Eq.~\eqref{eq:FDT}. The reservoir parameters are chosen as  $\mu\R = 0, \mu\L=\kB\bar{T}$, and the transmission parameters as $D_0 = 0.5, \Gamma = 0.5\kB\bar{T}, \epsilon_0 = 1.5 \kB\bar{T}$. 
    }
    \label{fig:FDB}
\end{figure}

To illustrate the fluctuation-dissipation bound~\eqref{eq:fluctuation-dissipation-bound}, and the resulting constraints~\eqref{eq:S-tot-bounds} for the total noise $S^I$, we show in Fig.~\ref{fig:FDB} their implications for a conductor with Lorentzian transmission,
\begin{equation}\label{eq:Lorentzian-transmission}
    D_\text{Lor}(E) = D_0 \frac{\Gamma^2}{\Gamma^2 + (E-\epsilon_0)^2},
\end{equation}
as well as for the complementary transmission function $1-D_\text{Lor}(E)$.
The constraints of Eq.~\eqref{eq:S-tot-bounds} define the grey excluded regions indicated in both panels,  in between which the noise $S^I$ (blue) for an arbitrary transmission $D(E)$ and in arbitrary out-of-equilibrium conditions can have intricate features. 
For the Lorentzian transmission, where the conductor is close to the tunneling regime, see panel (a), the total noise $S^I$ approaches the lower limit. 
By contrast, the opposite is true for the anti-Lorentzian transmission $[1-D_\text{Lor}(E)]$, where the conductor is close to the anti-tunneling regime and the noise approaches the upper limit,  see panel (b). 

Notably, the lower limit in Eq.~\eqref{eq:S-tot-bounds} can be negative, not providing useful constraints to the total noise $S^I$, which is always non-negative. Nonetheless, it is still relevant for the FDB for the (possibly negative) excess noise in Eq.~\eqref{eq:fluctuation-dissipation-bound}, as demonstrated by the green line in Fig.~\ref{fig:FDB}. In panel (b), it is obvious that the upper limit of Eq.~\eqref{eq:S-tot-bounds}, differing decisively from Eq.~\eqref{eq:Noise-Upper-Bound}, is not a strong constraint for the \emph{excess noise} (green line).
Furthermore, we see how Eq.~\eqref{eq:FDB-approximation} (dashed violet line) compares to the FDB in Eq.~\eqref{eq:fluctuation-dissipation-bound}: when the lower bound is negative ($P<0$), Eq.~\eqref{eq:FDB-approximation} is less constraining than Eq.~\eqref{eq:fluctuation-dissipation-bound}, but at larger $\Delta T$, Eq.~\eqref{eq:fluctuation-dissipation-bound} becomes a good approximation of the FDB.

As expected, the expression given by the FDT in Eq.~\eqref{eq:FDT} is reliable only for the Lorentzian transmission, which is close to the tunneling regime, and when the temperature bias $\Delta T$ is negligible. For sizable temperature biases, the expression of Eq.~\eqref{eq:FDT} is even excluded by the constraints for the noise~\eqref{eq:S-tot-bounds}.

\textit{Power fluctuation-dissipation bounds.---}
In the following, we consider thermoelectric engines, where we require the output power $P$ to be positive (the lower limit of~\eqref{eq:S-tot-bounds} is hence also positive).
Using Eq.~\eqref{eq:fluctuation-dissipation-bound} with $S^P = (\Delta\mu)^2S^I/q^2$, relating power and charge current noise and equivalently their noise components, we  show that the average output power is limited by the power fluctuations, through the following power FDB:
\begin{equation}\label{eq:Noise-Power-Bound}
    -\left(S^P-2\ThR^P\right)=\ThR^P - \ThL^P - \Ssh^P \geq  P \Delta\mu \tanh\left(\frac{\Delta\mu}{2\kB\Delta T}\right). 
\end{equation}
Specifically, large thermal fluctuations  $\ThR^P$ generated by the hot ``resource" contact allow for larger output power, whereas the thermal fluctuations $\ThL^P$ generated by the cold contact (absorbing power) diminish the maximum output power. This reflects the fact that, to increase the power output, one needs a large temperature bias $\Delta T$. 
Also the shot noise $\Ssh^P$, accounting for the ``friction" induced by the transmission $D(E)$, diminishes the possible power production. A favourable transmission $D(E)$ for a thermoelectric engine with large power output hence takes values $D(E)\in\{0,1\}$, as realized in step functions or box-car shaped transmissions~\cite{Whitney2014Apr, Whitney2015Mar}, making the shot noise vanish.

An alternative way to compare the output power with its fluctuations that has recently attracted a lot of attention is provided by the thermodynamic uncertainty relation (TUR), namely
\begin{equation}\label{eq:TUR}
    S^P \geq S^P_\text{TUR} \equiv 2\frac{\kB P^2}{\sigma}.    
\end{equation}
In addition to the power and its fluctuations, it explicitly contains the entropy production rate $\sigma$, in contrast to the bounds we develop in this Letter.
Since the TUR in Eq.~\eqref{eq:TUR} does not generally hold within scattering theory \cite{Brandner2018Mar, Agarwalla2018Oct, Liu2019Jun, Saryal2019Oct, Timpanaro2021Jun, Gerry2022Apr,Potanina2021Apr}, we find it instructive to compare it with the constraints arising from the power FDB of Eq.~\eqref{eq:Noise-Power-Bound}.
In fact, along with~\eqref{eq:Noise-Power-Bound}, we have constraints for the total power fluctuations at a given output power (and vice versa, see~\cite{supp}),
\begin{subequations}\label{eq:Total-Noise-Power-Bound}
    \begin{align}
        &S^P \geq P \Delta\mu \tanh\left(\frac{\Delta\mu}{2\kB\Delta T}\right),\label{eq:Noise-Power-lower-bound}\\
        &S^P\leq \frac{2\Delta\mu^2}{h}\kB\bar{T} + \left(P + \frac{\Delta\mu^2}{h}\right)\Delta\mu\tanh\left(\frac{\Delta\mu}{2\kB\Delta T}\right),
    \end{align}
\end{subequations}
which mirror the constraints on the total current fluctuations in Eq.~\eqref{eq:S-tot-bounds}.
\begin{figure}
    \centering
    \includegraphics{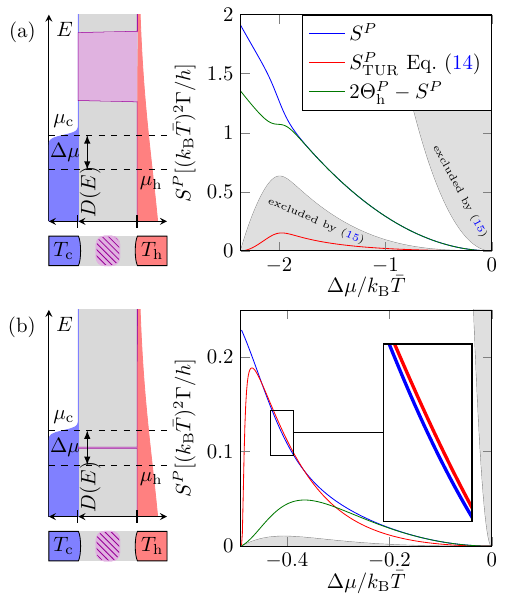}
    \caption{Comparison between  TUR [Eq.~\eqref{eq:TUR}] and the fluctuation-dissipation bound on the output power [Eq.~\eqref{eq:Total-Noise-Power-Bound}]  for a boxcar transmission $D_\text{box}(E)$ with (a) $\epsilon_0 = 3\kB\bar{T} = 3\Gamma$ and (b) $\epsilon_0 = 0.5\kB \bar{T} = 50\Gamma$. The average temperature $\Bar{T}$ is fixed and the temperature bias is  $\Delta T = 1.9\bar{T}$. }
    \label{fig:FDB-TUR}
\end{figure}
A comparison of these constraints for the full power fluctuations with the TUR are shown in Fig.~\ref{fig:FDB-TUR} for conductors with boxcar transmissions, namely
\begin{equation}
D_\text{box}(E) = \Big\{\begin{array}{ll}
    1 & \text{if}\quad E\in[\epsilon_0-\Gamma, \epsilon_0+\Gamma] \\
    0 & \text{otherwise}
\end{array},
\end{equation}
at different positions $\epsilon_0$ and half-width $\Gamma$. While Eqs.~\eqref{eq:Total-Noise-Power-Bound} provide both lower and upper constraints on the power fluctuations (grey regions), the TUR provides a lower limit only that is shown as a red line. 
Panel (a) demonstrates that the lower limit~\eqref{eq:Noise-Power-lower-bound} can provide a stronger constraint on the power fluctuations than the TUR, depending crucially on the range of $\Delta\mu$, the transmission $D(E)$ and the temperature bias $\Delta T$. 
This can in particular be the case close to the thermovoltage $\Delta\mu_{T}$, i.e. the electrochemical bias at which the power output vanishes $P(\Delta\mu_{T})=0$ (left boundary of panels (a) and (b) of Fig.~\ref{fig:FDB-TUR}, see also plots of the output power in the supplemental material~\cite{supp}). 
Indeed, since the thermovoltage is finite $\Delta\mu_{T}\neq 0$ in thermoelectric heat engines, when the engine operates close to $\Delta\mu_T$, the FDB of Eq.~\eqref{eq:Noise-Power-lower-bound} scales as $\mathcal{O}(hP/\Gamma\Delta \mu_T)$ whereas the TUR bound on the power fluctuations scales as $\mathcal{O}([hP/\Gamma\Delta\mu_T]^2)$, where $\Gamma$ is generally the energy-width of the transmission $D(E)$.

At low $\Gamma$ [see panel (b)] the TUR in Eq.~\eqref{eq:TUR} can be violated due to the break-down of detailed balance in this thermoelectric heat engine with high transmission in a chosen energy window. Importantly, the constraints of Eq.~\eqref{eq:Total-Noise-Power-Bound} emerging from the FDB still hold in this case, albeit they do not provide a tight limit in the chosen example.

In conclusion, this work provides an important step forward in the investigation of charge current fluctuations.
We establish universal constraints between the charge current and its noise, which we refer to as fluctuation-dissipation bounds and which hold for arbitrary noninteracting mesoscopic conductors  in general out-of-equilibrium conditions.
These bounds give important insights also for the out-of-equilibrium power noise of thermoelectric heat engines. Thereby, they provide constraints on their performance complementing the recently established thermodynamic uncertainty relations.



\textit{Acknowledgments.---}
Funding by the Knut and Alice Wallenberg foundation via the fellowship program (L.T. and J.S.) and by the Swedish Vetenskapsr\r{a}det via project number 2018-05061 (J.S.) is gratefully acknowledged.


\bibliography{refs.bib}

\end{document}


\title{Supplemental Material: Out-of-Equilibrium Fluctuation-Dissipation Bounds}

\author{Ludovico Tesser}
	\affiliation{Department of Microtechnology and Nanoscience (MC2), Chalmers University of Technology, S-412 96 G\"oteborg, Sweden}
	
	\author{Janine Splettstoesser}
	\affiliation{Department of Microtechnology and Nanoscience (MC2), Chalmers University of Technology, S-412 96 G\"oteborg, Sweden}
	
\date{\today}
\maketitle

\numberwithin{equation}{section}
\numberwithin{figure}{section}
\renewcommand{\thefigure}{\thesection\arabic{figure}}
\renewcommand{\theequation}{\thesection\arabic{equation}}
\renewcommand{\thesection}{\Alph{section}}
\section{Fluctuation dissipation theorems: Proofs of Eqs.~(\ref{eq:FDT}, \ref{eq:FDT2})}\label{app:FDT-proofs}
In this section we provide simple proofs of the out-of-equilibrium FDTs of Eqs.~(\ref{eq:FDT}, \ref{eq:FDT2}) within the framework of scattering theory.
In the tunneling regime, in which $D(E)\ll 1$, we can neglect the $D^2(E)$ term in the shot noise given in Eq.~\eqref{eq:noise-definitions} in the main text. In this weak-tunneling limit, the current noise reads
\begin{equation}
    S^I \approx \frac{q^2}{h} \!\!\int\!\! dE D(E)[\fL[1-\fR] + \fR[1-\fL]].
\end{equation}
Interestingly, the current can be written in a similar way, namely
\begin{equation}
    I = \frac{q}{h}\!\!\int \!\!dED(E)[\fL[1-\fR] - \fR[1-\fL]].
\end{equation}
To compare the current and noise expressions we use the properties of the ratio between Fermi functions
\begin{equation}\label{app:FermiRatio}
\frac{\fL}{1-\fL}\frac{1-\fR}{\fR} = e^{-\beta\L(E-\mu\L)}e^{\beta\R(E-\mu\R)}.
\end{equation}
When the reservoirs' temperatures are equal $T\L=T\R\equiv T$, the dependence on the energy $E$ in Eq.~\eqref{app:FermiRatio} cancels out. This allows us to write the current and the noise as
\begin{eqnarray}
    I = \frac{q}{h}\!\!\int dE D(E)\fL[1-\fR][1-e^{\beta(\mu\R-\mu\L)}],\\
    S^I\approx \frac{q^2}{h}\!\!\int\!\!dED(E)\fL[1-\fR][1+e^{\beta(\mu\R-\mu\L)}],
\end{eqnarray}
which imply the fluctuation-dissipation theorem of Eq.~\eqref{eq:FDT}:
\begin{equation}
    \frac{S^I}{qI} = \frac{1+e^{\beta(\mu\R- \mu\L)}}{1-e^{\beta(\mu\R- \mu\L)}} = \coth\left(\frac{\mu\L-\mu\R}{2\kB T}\right).
\end{equation}
Interestingly, an anologous derivation to the one that we here demonstrated for a weakly transmitting conductor can be done for a strongly transmitting conductor with the transmission function $\tilde{D}(E) = 1-D(E)$ such that $\tilde{D}(E)\ll 1$. The corresponding current and noise components, indicated with a tilde, fulfill
\begin{subequations} \label{supp:eq:tilde-relations}
\begin{align}
    \tilde{I} = \frac{q}{h} (\mu\L-\mu\R) - I, \label{eq:app:tildeI}\\
    \tilde{\Theta}_\alpha^I = \frac{q^2\kB T_\alpha}{h} - \Theta_\alpha^I,\label{eq:app:tildeTheta}\\
    \tilde{S}_\text{sh}^I = \Ssh^I.\label{eq:app:tildeSsh}
\end{align}
\end{subequations}
These relations allow us to find a connection between the fluctuations and the current in the transparent transmission limit $[1-D(E)]\ll 1$. In fact, assuming $T\L=T\R\equiv T$, we can use Eq.~\eqref{eq:FDT} together with Eqs.~\eqref{supp:eq:tilde-relations} finding
\begin{equation}
    \frac{2q^2 \kB T}{h} + \Ssh^I - \ThL^I- \ThR^I = \left(\frac{q^2(\mu\R-\mu\L)}{h} + qI\right)\coth\left(\frac{\mu\R-\mu\L}{2\kB T}\right)
\end{equation}
for a conductor with the transmission $\tilde{D}(E)\ll1$. Rearranging the terms one reaches Eq.~\eqref{eq:FDT2}.

\section{Out-of-equilibrum Fluctuation-dissipation bounds: Proofs of Eqs.~(\ref{eq:fluctuation-dissipation-bound}, \ref{eq:Noise-Upper-Bound})}\label{app:FDB-proofs}

In this section we prove the out-of equilibrium fluctuation-dissipation bounds for the excess noises $S^I-2\ThR^I$ and $S^I-2\ThL^I$. The difference between the noise components in Eq.~\eqref{eq:fluctuation-dissipation-bound} reads
\begin{equation}
\begin{split}
    2\ThR^I -S^I = \ThR^I-\ThL^I - \Ssh^I &= \frac{q^2}{h}\!\!\int\!\!dE D^2(E)[\fL-\fR]^2+\frac{q^2}{h}\!\!\int\!\!dE D(E)[\fL-\fR][2\fR -1].
\end{split}
\end{equation}
The first integral is non-negative because its integrand is always non-negative. Additionally, assuming without loss of generality $T\R>T\L$, we find the inequalities
\begin{subequations}\label{eq:app:eps_inequality}
\begin{align}
    \int_{\eps}^\infty\!\! dED[f\L-f\R]f\R \geq f\R(\epsilon)\int_{\eps}^\infty\!\! dED[f\L-f\R]\\
    \int_{-\infty}^{\eps}\!\! dED[f\L-f\R]f\R \geq f\R(\epsilon)\int_{-\infty}^{\eps}\!\! dED[f\L-f\R]
\end{align}
\end{subequations}
where $\eps$ is defined through $f\L(\eps) = f\R(\eps)$ and reads $\eps = (\mu\L T\R - \mu\R T\L)/\Delta T$. This specific energy $\eps$ is the one at which particle exchange takes place reversibly. 
Combining these inequalities~\eqref{eq:app:eps_inequality} one finds
\begin{equation}
    2\ThR^I - S^I = \ThR^I-\ThL^I - \Ssh^I \geq q[2f\R(\eps)-1] I = qI \tanh\left(\frac12\frac{\Delta\mu}{\kB\Delta T}\right),
\end{equation}
which is equivalent to the lower fluctuation-dissipation bound of Eq.~\eqref{eq:fluctuation-dissipation-bound}.

In a similar fashion to Sec.~\ref{app:FDT-proofs}, we now consider Eq.~\eqref{eq:fluctuation-dissipation-bound} for the transmission  $\tilde{D}(E) = 1-D(E)$ and denote the corresponding currents and noise components with a tilde.
These quantities satisfy Eqs.~\eqref{supp:eq:tilde-relations} and thus we find
\begin{equation}
    \frac{q^2\kB\Delta T}{h} -\ThR^I +\ThL^I -\Ssh^I \geq -\left(\frac{q^2\Delta\mu}{h} + qI\right)\tanh\left(\frac12\frac{\Delta\mu}{\kB\Delta T}\right),
\end{equation}
which, once the terms are rearranged appropriately, coincides with the upper fluctuation-dissipation bound of Eq.~\eqref{eq:Noise-Upper-Bound}.

\section{Multi-terminal and multi-channel case: proof of Eq.~(\ref{eq:FDB_general})}

In this section, we extend the proof of Sec.~\ref{app:FDB-proofs} to the more general multi-terminal and multi-channel case.
When considering the multi-channel case it is convenient to decompose the scattering matrix into subblocks $t_{\alpha\beta}$ as
\begin{equation}
    s \equiv \begin{pmatrix}
        t_{11} & t_{12} & \cdots \\
        t_{21} & t_{22} & \cdots \\
        \vdots & \vdots & \ddots
    \end{pmatrix}
\end{equation}
where $t_{\alpha\beta}$ are matrices of dimension $N_\alpha\times N_\beta$, with $N_\alpha$ the number of channels into reservoir $\alpha$. These matrices describe transport between the $N_\beta$ channels of reservoir $\beta$ to the $N_\alpha$ channels of reservoir $\alpha$. Note that the scattering matrix, and therefore the $t_{\alpha\beta}$ matrices, are energy-dependent. However, to ease the notation, we do in this section not write the energy dependence of any function unless we evaluate it at a specific energy.
Since the scattering matrix is unitary, i.e. $s^\dagger s = s s^\dagger = \mathbb{I}$, the transmission matrices $t_{\alpha\beta}$ satisfy
\begin{equation}
    \sum_{\beta} t_{\alpha\beta}t_{\alpha\beta}^\dagger = \mathbb{I}_\alpha\quad\Rightarrow\quad \text{Tr}\left\{\sum_\beta t_{\alpha\beta}t_{\alpha\beta}^\dagger\right\} = N_\alpha.
\end{equation}
The zero-frequency autocorrelator is written in terms of the transmission matrices as~\cite{Blanter2000Sep}
\begin{equation}
    S_{\alpha\alpha}^I = \frac{q^2}{h}\int dE\left[\sum_{\beta,\gamma\neq \alpha}\text{Tr}\left\{t_{\alpha\beta}t_{\alpha\beta}^\dagger t_{\alpha\gamma}t_{\alpha\gamma}^\dagger\right\} (f_\beta - f_\alpha)(f_\alpha - f_\gamma) + \sum_{\beta\neq\alpha}\text{Tr}\left\{t_{\alpha\beta}t_{\alpha\beta}^\dagger\right\}\left(f_\alpha(1-f_\beta) + f_\beta(1-f_\alpha)\right)\right].
\end{equation}
The first contribution corresponds to the quantum component of the noise $S_{\alpha\alpha, \text{qu}}^I$, whereas the second one is the classical component $S_{\alpha\alpha, \text{cl}}^I$.
Note that, by introducing the Hermitian matrix $M_\alpha \equiv \sum_{\beta\neq\alpha} t_{\alpha\beta}t_{\alpha\beta}^\dagger (f_\beta-f_\alpha) = M_\alpha^\dagger$, the quantum noise component is written as
\begin{equation}
    S_{\alpha\alpha, \text{qu}}^I = -\frac{q^2}{h}\int dE \text{Tr}\left\{M_\alpha^2\right\}\leq 0.
\end{equation}
This expression is negative or zero, since $M_\alpha$ is Hermitian. Its eigenvalues are therefore real, implying that the eigenvalues of $M_\alpha^2$ are non-negative.
In particular, the quantum noise component vanishes at equilibrium: when $f_\beta = f_\alpha$ $\forall\beta$,  $M_\alpha = 0$.

Furthermore, by introducing the eigenvalues $D_{\alpha\beta, i}\in[0,1]$ of the matrices $t_{\alpha\beta}t_{\alpha\beta}^\dagger$ the average current into reservoir $\alpha$ and the classical noise component read
\begin{equation}
    I_\alpha = \frac{q}{h} \int dE \sum_{\beta_\neq\alpha}\sum_{i=1}^{N_\alpha} D_{\alpha\beta, i}(f_\alpha - f_\beta),
\end{equation}
\begin{equation}
    S_{\alpha\alpha, \text{cl}}^I = \frac{q^2}{h}\int dE \sum_{\beta\neq \alpha}\sum_{i=1}^{N_\alpha}D_{\alpha\beta, i}\left[f_\beta(1-f_\alpha)+ f_\alpha(1-f_\beta)\right].
\end{equation}
Comparing the out-of-equilibrium noise with the equilibrium one, in which $f_\beta = f_\alpha$ $\forall\beta$, we find
\begin{equation}\label{eq:app:excess_multi}
    S_{\alpha\alpha}^I - S_{\alpha\alpha, \text{eq}}^I = S_{\alpha\alpha, \text{qu}}^I  + \frac{q^2}{h}\int dE\sum_{\beta\neq \alpha}\sum_{i=1}^{N_\alpha}D_{\alpha\beta, i}\left[f_\beta(1-f_\alpha)+ f_\alpha(1-f_\beta) - 2 f_\alpha(1-f_\alpha)\right].
\end{equation}
The integral in the excess out-of-equilibrium noise of Eq.~\eqref{eq:app:excess_multi} can be written as
\begin{equation}\label{eq:app:multi_int}
    \frac{q^2}{h}\int dE\sum_{\beta\neq \alpha}\sum_{i=1}^{N_\alpha}D_{\alpha\beta, i}(f_\alpha-f_\beta)(2f_\alpha -1)= -qI_\alpha +  2\frac{q^2}{h}\int dE\sum_{\beta\neq \alpha}\sum_{i=1}^{N_\alpha}D_{\alpha\beta, i}(f_\alpha-f_\beta)f_\alpha.
\end{equation}
Calling $\epsilon_{\alpha\beta}$ the energy at which $f_\alpha(\epsilon_{\alpha\beta}) = f_\beta (\epsilon_{\alpha\beta})$, i.e. the energy at which the Fermi distributions of reservoirs $\alpha$ and $\beta$ cross, we can split the integral on the right hand side of Eq.~\eqref{eq:app:multi_int} according to the sign of $f_\alpha-f_\beta$ as
\begin{equation}\label{eq:app:multi_int_split}
    2\frac{q^2}{h}\int dE\sum_{\beta\neq \alpha}\sum_{i=1}^{N_\alpha}D_{\alpha\beta, i}(f_\alpha-f_\beta)f_\alpha = 2\frac{q^2}{h}\sum_{\beta\neq \alpha}\int_{-\infty}^{\epsilon_{\alpha\beta}} \!\!\!\!dE\sum_{i=1}^{N_\alpha}D_{\alpha\beta, i}(f_\alpha-f_\beta)f_\alpha + 2\frac{q^2}{h}\sum_{\beta\neq \alpha}\int_{\epsilon_{\alpha\beta}}^\infty \!\!\!\!dE\sum_{i=1}^{N_\alpha}D_{\alpha\beta, i}(f_\alpha-f_\beta)f_\alpha.
\end{equation}
We now assume that $T_\alpha\equiv T_\h$ is the \textit{largest} temperature, such that the distribution $f_\alpha\equiv f_\h$ has the smoothest transition from 1 to 0 compared to all other $f_{\beta\neq\h}$, and use that $f_\h$ is a decreasing function of energy. Then, the two integrals of Eq.~\eqref{eq:app:multi_int_split} satisfy the following inequalities
\begin{subequations}\label{supp:multi-terminal-channel-inequalities}
    \begin{align}
        2\frac{q^2}{h}\sum_{\beta\neq \h}\int_{-\infty}^{\epsilon_{\h\beta}} \!\!\!\!dE\sum_{i=1}^{N_\h}D_{\h\beta, i}(f_\h-f_\beta)f_\h   &\leq    2\frac{q^2}{h}\sum_{\beta\neq \h}f_\h(\epsilon_{\h\beta})\int_{-\infty}^{\epsilon_{\h\beta}} \!\!\!\!dE\sum_{i=1}^{N_\h}D_{\h\beta, i}(f_\h-f_\beta), \\
         2\frac{q^2}{h}\sum_{\beta\neq \h}\int_{\epsilon_{\h\beta}}^\infty \!\!\!\!dE\sum_{i=1}^{N_\h}D_{\h\beta, i}(f_\h-f_\beta)f_\h   &\leq  2\frac{q^2}{h}\sum_{\beta\neq \h}f_\h(\epsilon_{\h\beta})\int_{\epsilon_{\h\beta}}^\infty \!\!\!\!dE\sum_{i=1}^{N_\h}D_{\h\beta, i}(f_\h-f_\beta).
    \end{align}
\end{subequations}
Combining them, we find the following bound on the excess noise of the \textit{hottest} contact
\begin{subequations}
    \begin{align}
        S_{\h\h}^I - S_{\h\h, \text{eq}}^I &\leq S_{\h\h, \text{qu}}^I -qI_\h + 2\frac{q^2}{h}\sum_{\beta\neq \h}f_\h(\epsilon_{\h\beta})\int dE\sum_{i=1}^{N_\h}D_{\h\beta, i}(f_\h-f_\beta) \label{supp:eq:FDB-multi-terminal-neq} \\
        S_{\h\h}^I - S_{\h\h, \text{eq}}^I &\leq \frac{q^2}{h}\sum_{\beta\neq\h}\tanh\left(\frac12\frac{\mu_\h-\mu_\beta}{\kB(T_\h-T_\beta)}\right)\int dE \sum_{i=1}^{N_\h}D_{\h\beta, i}(f_\h-f_\beta),
    \end{align}
\end{subequations}
which reduces to Eq.~\eqref{eq:fluctuation-dissipation-bound} when there are only two reservoirs and a single channel.
Note that Eq.~\eqref{supp:eq:FDB-multi-terminal-neq} also holds for nonthermal distributions $f_\beta$ as long as $f_\h(E)-f_\beta(E)=0$ has only one solution for $\h\neq\beta$, where the ``hottest" distribution $f_\h$, i.e. the distribution  that satisfies $f_\h(E)>f_\beta(E)$ for all $\beta\neq\h$ at a sufficiently large $E$, is decreasing.

If we instead consider the excess noise in the \textit{coldest} reservoir $T_\alpha \equiv T_\c$, the inequalities in Eq.~\eqref{supp:multi-terminal-channel-inequalities}
 change direction, and we find a lower bound on the excess classical noise of the coldest contact
 \begin{equation}
     S^I_{\c\c} -S_{\c\c, \text{qu}}^I -S_{\c\c, \text{eq}}^I =S^I_{\c\c, \text{cl}} -S_{\c\c, \text{eq}}^I \geq  \frac{q^2}{h}\sum_{\beta\neq\c}\tanh\left(\frac12\frac{\mu_\c-\mu_\beta}{\kB(T_\c-T_\beta)}\right)\int dE \sum_{i=1}^{N_\c}D_{\c\beta, i}(f_\c-f_\beta).
 \end{equation}
Importantly, since these inequalities provide a generalization of the FDBs~(\ref{eq:fluctuation-dissipation-bound}) and (\ref{eq:Noise-Upper-Bound}) of the main text to  multi-terminal multi-channel conductors, this also implies that the FDBs are valid for an even broader range of out-of-equilibrium conditions. Situations like spin biases or out-of-equilibrium situations between different orbital degrees of freedom can be implemented via identifying them as a bias between effectively multiple terminals. 

\section{Constraints on power production}

\begin{figure}[h]
    \centering
    \includegraphics[scale=1]{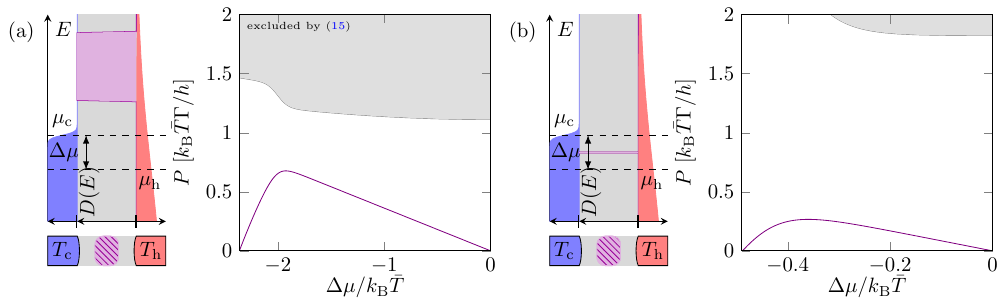}
    \caption{Output power $P$ and its constraints emerging from Eq.~\eqref{eq:Total-Noise-Power-Bound}  for a boxcar transmission $D_\text{box}(E)$ with (a) $\epsilon_0 = 3\kB\bar{T} = 3\Gamma$ and (b) $\epsilon_0 = 0.5\kB \bar{T} = 50\Gamma$. The average temperature $\Bar{T}$ is fixed and the temperature bias is  $\Delta T = 1.9\bar{T}$.}
    \label{supp:fig:power}
\end{figure}

The constraints on the power fluctuations in Eq.~\eqref{eq:Total-Noise-Power-Bound} can also be cast as constraints on the output power $P$.
Specifically,
\begin{equation}
    \frac{S^P}{\Delta\mu\tanh\left(\frac{\Delta\mu}{2\kB\Delta T}\right)} -\frac{\Delta\mu^2}{h}\left[1+ \frac{2\kB\bar{T}}{\Delta\mu\tanh\left(\frac{\Delta\mu}{2\kB\Delta T}\right)}\right] \leq P \leq \frac{S^P}{\Delta\mu\tanh\left(\frac{\Delta\mu}{2\kB\Delta T}\right)}.
\end{equation}
These are shown in Fig.~\ref{supp:fig:power} for the same setups used in Fig.~\ref{fig:FDB-TUR} in the main text. Here, only the upper bound is visible because, in the selected examples, the lower bound is negative while we are interested in the thermoelectric heat engine regime, i.e. $P>0$.

\bibliography{refs.bib}